\documentclass[finalnew]{agujournal2019}
\usepackage{url} %this package should fix any errors with URLs in refs.
\usepackage{lineno}
\usepackage[finalnew]{trackchanges}
\usepackage{soul}
\usepackage{amsmath}
\usepackage{amsfonts}
\usepackage{amssymb}
\graphicspath{{figs/}}

\draftfalse

\journalname{Geophysical Research Letters}

\renewcommand{\vec}{\mathbf}

\newcommand{\B}{\vec{B}}
\newcommand{\vel}{\vec{u}}

\newcommand{\pos}{\vec{r}}

\newcommand{\ez}{\vec{1}_z}

\newcommand{\rotax}{\vec{1}_z}

\newcommand{\Le}{\mathrm{Le}}
\newcommand{\Lu}{\mathrm{Lu}}
\newcommand{\Pm}{\mathrm{Pm}}

\newcommand{\dV}{\mathrm{d}V}

\newcommand{\bcdot}{\boldsymbol{\cdot}}
\newcommand{\divergence}{\boldsymbol{\nabla}\bcdot}
\newcommand{\curl}{\boldsymbol{\nabla}\times}

\newcommand{\vintin}[1]{\int_\mathcal{V} #1\,\dV}

\newcommand{\R}{\mathbb{R}}
\newcommand{\Cn}{\mathbb{C}}

\newcommand{\ekmr}{E_\mathrm{kin}/E_\mathrm{mag}}

\begin{document}

\title{Fast Quasi-Geostrophic Magneto-Coriolis Modes in the Earth's core}

\authors{F. Gerick\affil{1,2}, D. Jault\affil{1}, J. Noir\affil{2}}

\affiliation{1}{CNRS, ISTerre, University of Grenoble Alpes, Grenoble, France}
\affiliation{2}{Institute of Geophysics, ETH Zurich, Zurich, Switzerland}

\correspondingauthor{Felix Gerick}{felix.gerick@univ-grenoble-alpes.fr}

\begin{keypoints}
\item Magneto-Coriolis modes of periods close to torsional Alfv\'en modes could be present in Earth's core model without stratification.
\item The magnetic field changes of such modes show properties similar to geomagnetic observations, with fast changes localized near the equator.
\item Our model could allow core-flow inversions from geomagnetic field data to the flow and simultaneously the magnetic field within the core.
\end{keypoints}

\begin{abstract}
      Fast changes of Earth's magnetic field could be explained by inviscid and diffusion-less quasi-geostrophic (QG) Magneto-Coriolis modes.
      We present a hybrid QG model with columnar flows and three-dimensional magnetic fields and find modes with \change{frequencies}{periods} of a few years at parameters relevant to Earth's core.
      For the simple poloidal magnetic field that we consider here they show a localization of kinetic and magnetic energy in the equatorial region.
      This \change{focusing}{concentration of energy} near the equator and the high frequency make them a plausible mechanism to explain similar features observed in recent geomagnetic field observations.
      Our model potentially opens a way to probe the otherwise inaccessible magnetic field structure in the Earth's outer core. 
\end{abstract}

\section{Introduction}

\citeA{hide_free_1966} proposed that temporal changes of Earth's magnetic field, called secular variations (SV), could originate from linear modes present in the Earth's liquid outer core.
These modes are separated into modes dominated by a balance of magnetic, Coriolis and pressure forces, known as Magneto-Coriolis modes (MCM), and modes dominated by inertial, Coriolis and pressure forces.
The latter are often referred to as quasi-geostrophic (QG) inertial modes, or Rossby modes (RM). Torsional Alfv\'en modes (TM), consisting of geostrophic motions \cite{braginsky_torsional_1970}, complete the set of incompressible and diffusion-less magnetohydrodynamic (MHD) modes. They obey a balance between inertia and the magnetic force.
In this study, we use a reduced model, based on the QG assumption for the velocity and a three-dimensional (\change{3D}{3-D}) magnetic field that is compatible with an insulating mantle, to investigate the SV associated \change{to}{with} such QG modes, with a focus on MCM, in the Earth's core.

Monitoring the radial magnetic field component at the core-mantle boundary (CMB) is the main way of probing flows in the liquid outer core of Earth. 
A large number of studies are concerned with the inversion of the downward projected geomagnetic field to flows in the outer core, a process referred to as core-flow inversion \cite<see>[for a review]{holme_804_2015}.
The most commonly applied core-flow inversions are based on geostrophic flows tangential to the CMB \cite{le_mouel_outercore_1984,chulliat_local_2000}. 
These inversions give the flow field local to the CMB. 
Several works have then inferred core dynamics from the inverted flow field at the CMB.
\citeA{zatman_torsional_1997} and more recently \citeA{gillet_fast_2010} have inverted these surface flows to the mean radial magnetic field component within Earth's core through TM.
\citeA{buffett_geomagnetic_2014} correlated Magneto-Archimedes-Coriolis (MAC) waves in a stably-stratified layer at the top of the outer core with the inferred surface flows. 

Our model potentially serves as a new forward model to invert geomagnetic field observations.
In this approach a reduced set of MHD equations is solved in the bulk of the fluid.
It is based on the QG assumption for the velocity, where a balance between Coriolis and pressure gradient forces is dominant, while allowing linear axial dependence of the flow field. 
This assumption is appropriate to investigate fluids under rapid rotation at time scales much larger than the rotation period. 
Different studies have shown that a large part of the inferred surface core flow is equatorially symmetric and may account for a large part of the observed secular variations \cite{gillet_ensemble_2009,gillet_rationale_2011}.
Additionally, recent \change{3D}{3-D} high resolution numerical simulations revealed a largely columnar flow structure, in agreement with the QG assumption \cite<e.g.>[]{schaeffer_turbulent_2017}.
\add{To establish and maintain such a columnar flow within the Earth's core different mechanisms have been proposed, e.g. 3-D inertial-Alfv\'en waves that transport energy along the rotation axis} \cite{bardsley_inertialalfven_2016}\add{.
Such built up of columns occurs on diurnal periods and are thus not captured in the QG model, where a columnar structure is assumed to be already established.}
Previously, consistently derived QG models that include a magnetic field \change{where}{were} limited to magnetic fields that treated the CMB as a perfectly conducting boundary \cite{busse_generation_1976, canet_hydromagnetic_2014,labbe_magnetostrophic_2015,gerick_pressure_2020}.
For such magnetic fields the radial component at the boundary must vanish, rendering them unsuitable to associate core flows with magnetic field imprints at the CMB.
There have been approaches to combine a QG model with a magnetic field that has a non-zero radial magnetic field component at the surface, but they rely on the neglect of surface terms in the induction equation that is difficult to justify \cite{canet_forward_2009,maffei_kinematic_2017}.
Here we present a hybrid model that combines QG velocities with a \change{3D}{3-D} insulating magnetic field. 
We follow the approach presented in \citeA{gerick_pressure_2020} with a new basis for the magnetic field that satisfies the insulating boundary condition at the CMB.
Both the QG velocity and the magnetic field basis vectors are expressed in Cartesian polynomials. This methodology has been fruitful to model modes and instabilities in rapidly rotating ellipsoids \cite{vantieghem_inertial_2014,vidal_inviscid_2017,vidal_fossil_2019,vidal_compressible_2020}.
This Cartesian presentation of the basis vectors is particularly useful for the Galerkin approach used here, due to the easy integration of Cartesian monomials over the volume \cite{lebovitz_stability_1989}.
We derive a basis for the magnetic field in Cartesian polynomials, that exploits the properties of spherical harmonics.

\section{A Hybrid Quasi-Geostrophic Model and Columnar Modes}
\subsection{Magnetohydrodynamic Equations}

The equations governing the incompressible flow $\vel$ and the magnetic field $\B$ in a rapidly rotating planetary core of volume $\mathcal{V}$\add{, here assumed to be a full sphere without an inner core,} are given in non-dimensionalized form by 
\begin{subequations}
    \label{eq:goveqadimNL}
\begin{linenomath*}
\begin{align}
            \label{eq:momeqnondim}
        \frac{\partial \vel}{\partial t} + (\vel\boldsymbol{\cdot}\boldsymbol{\nabla})\, \vel =& -\frac{2}{\Le} \, \rotax\times\vel  -\nabla p + \frac{\Pm}{\Lu}\, \boldsymbol{\nabla}^2\vel+ (\boldsymbol{\nabla}\times\B)\times\B,\\
        \frac{\partial \B}{\partial t} =& \boldsymbol{\nabla} \times\left(\vel\times\B\right)+\frac{1}{\Lu} \, \boldsymbol{\nabla}^2\B.
\end{align}
\end{linenomath*}
\end{subequations}
The non-dimensional Lehnert, Lundquist and magnetic Prandtl number are given by
\begin{linenomath*}
 \begin{equation}
    \Le=\frac{B_0}{\Omega R_0 \sqrt{\mu_0\rho}}, \quad \Lu = \frac{R_0B_0}{\eta\sqrt{\mu_0\rho}}, \quad \Pm = \frac{\nu}{\eta},
\end{equation}
\end{linenomath*}
with $\vec{\Omega}=\Omega\rotax$ the rotation vector, $\rho$ the fluid density, $p$ the reduced pressure, $\nu$ the kinematic viscosity, $\mu_0$ the permeability of vacuum, $\eta$ the magnetic diffusivity, $R_0$ the core radius and $B_0$ the characteristic strength of the magnetic field. 
The characteristic time scale is the Alfv\'en time scale $T_A = R_0/u_A$, where $u_A=B_0/\sqrt{\rho \mu_0}$ is the characteristic Alfv\'en velocity. 
Equations \eqref{eq:goveqadimNL} are subject to the non-slip boundary condition $\vel=\vec{0}$ and the continuity of the magnetic field across the boundary $[\B]=\vec{0}$, where $[\cdot]$ denotes a jump.

For parameters relevant for Earth's core, $\Le \sim 10^{-4}$, $\Lu \sim 10^5$ and $\Pm \sim 10^{-6}$ \cite{wijs_viscosity_1998, gillet_fast_2010, pozzo_thermal_2014}. 
Thus, if we additionally consider time scales on the order of $T_A$, it is appropriate to neglect viscous and diffusive effects in the bulk. 
In the next step, since we are interested in the linear response of the system, the velocity and magnetic field are perturbed around a background state with no motion and steady magnetic field $\B_0$.
In the Earth's core, the characteristic mean velocity field is thought to be negligible compared to the Alfv\'en wave velocity \cite{gillet_planetary_2015,barenzung_modeling_2018}.
Hence, the equations describing the evolution of the velocity and magnetic perturbations $[\tilde\vel,\tilde\B]$ are given by
\begin{subequations}
\label{eq:goveqadimLN}
\begin{linenomath*}
\begin{align}
    \frac{\partial \tilde\vel}{\partial t} + \frac{2}{\Le}\, \rotax\times\tilde\vel =& -\nabla p + (\boldsymbol{\nabla}\times\B_0)\times\tilde\B+ (\boldsymbol{\nabla}\times\tilde\B)\times\B_0, \label{eq:linmom}\\
    \frac{\partial \tilde\B}{\partial t} =& \boldsymbol{\nabla}\times\left(\tilde\vel\times\B_0\right).\label{eq:linind}
\end{align}
\end{linenomath*}
\end{subequations}
In the limit $\Pm\rightarrow 0$ the boundary conditions on the velocity reduces to the non-penetration condition $\vel\bcdot\vec{n}=0$\add{, with $\vec{n}$ the vector normal to the boundary,} and the magnetic boundary condition is not modified \cite{stewartson_dispersion_1957,hide_hydromagnetic_1972}. 
\change{By doing so we assume that the motions that we are investigating are able to eliminate any current layer on the fluid surface, in contrast to what has been initially assumed for the equation of TM}{Previous studies allowed for a jump in the tangential component of the magnetic field across a diffusive boundary layer}  \cite{braginsky_torsional_1970,jault_809_2015}.
\add{Here, we assume that the motions that we are investigating are able to eliminate any current layer on the fluid surface.}

\subsection{Quasi-Geostrophic Velocity Basis}

Assuming that the equatorial components of the velocity are independent of the coordinate $z$ along the rotation axis; the non-penetration boundary condition, $\vel\cdot\vec{n}=0$ on \add{the core-mantle boundary} $\partial\mathcal{V}$ holds and the flow is incompressible, $\divergence\vel=0$, the quasi-geostrophic (QG) velocity takes the form \cite{amit_helical_2004, schaeffer_quasigeostrophic_2005, bardsley_could_2018}
\begin{linenomath*}
 \begin{equation}
    \vel=\nabla\psi\times\nabla\left(\frac{z}{h}\right), \label{eq:qgvel}
\end{equation}
\end{linenomath*}
with $h$ the \add{half} height of the fluid column and $\psi$ a scalar stream function depending only on the horizontal coordinates.

In Cartesian coordinates the stream function can be expressed as \cite{maffei_propagation_2016,gerick_pressure_2020}
\begin{linenomath*}
 \begin{equation}
  \psi_i = h^3 \, \Pi_i,
\end{equation}
\end{linenomath*}
with $\Pi_i$ \add{being} a monomial in \add{the equatorial Cartesian coordinates} $x$ and $y$ of degree $N$, so that $i\in [0,N_2]$ with $N_2=N(N+1)/2$. The QG basis vectors $\vel_i$ are given by
\begin{linenomath*}
 \begin{equation}
  \vel_i = h^2\nabla\Pi_i\times\ez + 3\Pi_i\nabla G\times\ez-z\nabla\Pi_i\times\nabla G,
  \label{eq:qg-basis}
\end{equation}
\end{linenomath*}
with $\nabla G=h\nabla h = -x\vec{1}_x-y\vec{1}_y$.

\subsection{Magnetic Field Basis\label{sec:magbasis}}
In this section we present a set of basis vectors for the 3-D magnetic field, satisfying insulating boundary conditions at the CMB. 
\add{Unlike in classical geodynamo simulations, where the boundary condition is enforced at each time step of the forward iteration, the boundary condition is included in the basis elements} \cite{zhang_hydromagnetic_1995, li_optimal_2010, chen_optimal_2018}\add{. The detailed derivation of such a basis is given in} \ref{a:magbasis}.

We write the magnetic field $\B$ in the toroidal-poloidal expansion, so that
\begin{linenomath*}
 \begin{equation}
    \vec{B} = \B_t + \B_p =  \curl T\pos + \curl\curl P\pos.
\end{equation}
\end{linenomath*}

The toroidal and poloidal scalars are written for each spherical harmonic degree $l$, order $m$ and radial degree $n$, so that
\begin{linenomath*}
\begin{align}
    T_{lmn} &= (1-r^2)r^{2n}R_l^m,\\
    P_{lmn} &= -r^{2(n+1)}\frac{R_l^m}{2(n+1)(2(l+n)+3)}.
\end{align}
\end{linenomath*}
with $R_l^m = r^lY_l^m(\theta,\phi)$ the solid spherical harmonics. We have $|m|\leq l$ and $l \in [1,N]$, $n\in[0,(N-l)/2\big\rfloor$ for the toroidal basis and $l \in [0,N-1]$, $n\in[0,(N+1-l)/2-1\big\rfloor$ for the poloidal basis, resulting in a total of $N_3=\frac{1}{6}N(N+1)(2N+7)$ basis vectors.
The toroidal part of $\B$ is given by
\begin{linenomath*}
 \begin{equation}
    \B_{t,lmn} = \curl T_{lmn} \pos. \label{eq:btor}
\end{equation}
\end{linenomath*}
The poloidal magnetic field has to satisfy the continuity at the core-mantle boundary $\partial\mathcal{V}$, so that ${\nabla \Phi^i + \B_p = \nabla \Phi^e}$,
with $\Phi^i$ and $\Phi^e$ the interior and exterior potential field, respectively. The exterior potential field must vanish at infinity, if the source of the magnetic field lies within the interior.
The poloidal basis vectors are thus given by
\begin{linenomath*}
 \begin{equation}
    \B_{p,lmn} = \curl\curl P_{lmn}\pos + \nabla \Phi^i_{lmn},\label{eq:bpol}
\end{equation}
\end{linenomath*}
with $\Phi^i_{lmn}=-\frac{(l + 1)}{(2l + 1)(2n + 2)}r^lY_{l}^m$.

Together with the toroidal component \eqref{eq:btor}, this basis can be transformed to Cartesian coordinates by representing the spherical harmonics in terms of unit Cartesian coordinates, as presented in \ref{a:csh}.
We choose the Schmidt-semi normalization for the spherical harmonics, but any other may be chosen, as we normalize the basis vectors afterwards, so that $\vintin{\B_i\bcdot\B_i} = 1$.

\subsection{Projection Method}
\change{A}{We introduce a} hybrid quasi-geostrophic (QG) model with QG velocities and 3-D magnetic field \remove{is introduced}, following \citeA{gerick_pressure_2020}. \remove{We project the linearized momentum equation (3a) onto a QG basis u' of the form of (4). Similarly, the linearized induction equation (3b) is projected onto a basis B' of 3-D magnetic fields of the form (13) and (27).} The linearized momentum equation \eqref{eq:linmom} and induction equation \eqref{eq:linind} are projected onto a QG basis $\vel'$ of the form of \eqref{eq:qgvel} and a 3-D magnetic field basis $\B'$ of the form \eqref{eq:btor} and \eqref{eq:bpol}, respectively. This method is essentially a variational approach, which consists in finding solutions $[\tilde\vel,\tilde\B]$ satisfying
\begin{subequations}
\label{eq:varapproach}
\begin{linenomath*}
\begin{align}
    \begin{split}
    \vintin{\vel'\bcdot \frac{\partial \tilde\vel}{\partial t}} =& -\vintin{\vel'\bcdot\left(\frac{2}{\Le}\rotax\times\tilde\vel+\nabla p\right)}\\
    &+\vintin{\vel'\bcdot\left( \left(\boldsymbol{\nabla}\times\B_0 \right)\times\tilde\B+ (\boldsymbol{\nabla}\times\tilde\B)\times\B_0\right)} \end{split}\label{eq:varapproach_a}\quad\forall\,\vel'\\
    \vintin{\B'\bcdot \frac{\partial \B}{\partial t}} =& \vintin{\B'\bcdot \curl\left(\tilde\vel\times\B_0\right)}\quad \forall\,\B'
\end{align}
\end{linenomath*}
\end{subequations}

This set of equations may be reduced to a scalar evolution equation for the stream function $\psi$ accompanied by the \change{3D}{3-D} induction equation, as shown in \citeA<equation \change{(30)}{(48)} of>[]{gerick_pressure_2020}.
\add{This hybrid model has been verified against a fully 3-D model at moderate polynomial truncation} \cite<see also>[]{gerick_pressure_2020}.

When replacing the test functions $\vel'$ in \eqref{eq:varapproach_a} by the subset of purely geostrophic velocities $\vel_G = u_G(s)\vec{1}_\phi$
we obtain the equation for the diffusion-less torsional Alfv\'en modes (TM) initially discovered by \citeA{braginsky_torsional_1970}.
The one-dimensional (\change{1D}{1-D}) TM equation \change{reads}{is written}
\begin{linenomath*}
 \begin{equation}
    s^3h\frac{\partial^2\xi}{\partial t^2} = \frac{\partial}{\partial s} \left(s^3h v_A^2\frac{\partial \xi}{\partial s}\right) \label{eq:1dtm}
\end{equation}
\end{linenomath*}
with $\xi=u_G(s)/s$ and the mean squared cylindrical Alfv\'en velocity
\begin{linenomath*}
 \begin{equation}
    v_A^2(s) = \frac{1}{4\pi s h} \oint\int  (\B_0\bcdot\vec{1}_s)^2s\,\mathrm{d}z\mathrm{d}\phi.
\end{equation}
\end{linenomath*}

For more details on the derivation we refer the reader to \citeA{jault_electromagnetic_2003}.
Equation \eqref{eq:1dtm} diverges near the equator as $s\rightarrow 1$, but solutions exist, if $v_A^2(1)\neq 0$ \cite{maffei_propagation_2016}.
Since we can compute solutions to the \change{1D}{1-D} equation for a given background magnetic field satisfying these conditions, TM suit well as a benchmark of our hybrid QG model capable of capturing TM.

\subsection{Quasi-geostrophic Inertial and Magneto-Coriolis Modes}

\citeA{lehnert_magnetohydrodynamic_1954} introduced two distinct families of MHD modes as solutions to the linearized MHD equations \eqref{eq:goveqadimLN}, namely slow MCM and fast, slightly modified RM.
The phase velocity is prograde for slightly modified RM and retrograde for MCM.
\citeA{malkus_hydromagnetic_1967} has shown that for an idealized magnetic field $\B_{0,M} = s\vec{1}_\phi$ of uniform current density analytical solutions exists for these two mode families. 
The dispersion relations are given by \cite{labbe_magnetostrophic_2015}
\begin{linenomath*}
 \begin{equation}
    \omega^\pm_{n,m} = \frac{1}{2\Le}\lambda_{n,m}\left( 1\pm\left(1+\frac{4\Le^2 m(m{-}\lambda_{n,m})}{\lambda_{n,m}^2}\right)^{1/2} \right),\label{eq:malkus}
\end{equation}
\end{linenomath*}
with $\omega^+_{n,m}$, $\omega^-_{n,m}$ and $\lambda_{n,m}$ the frequencies of the slightly modified RM, MCM and hydrodynamic (HD) inertial modes of azimuthal wave number $m$ and radial scale \change{$N$}{$n$}, respectively. 
The dispersion relation shows that the difference between $\omega^+_{n,m}$ and $\lambda_{n,m}/\Le$ is small, if $\Le\ll 1$.
The frequencies $\lambda_{n,m}$ are scaled by the rotation frequency and can be obtained as solutions to a univariate polynomial in the sphere \cite{zhang_inertial_2001}. An approximate value for the equatorially symmetric inertial modes is given by \cite{zhang_inertial_2001}
\begin{linenomath*}
 \begin{equation}
    \lambda_{n,m} \approx -\frac{2}{m+2}\left( \left(1+\frac{m(m+2)}{n(2n+2m+1)}\right)^{1/2}-1 \right).\label{eq:zhangrossby}
\end{equation}
\end{linenomath*}

As the magnetic field perturbations in the Malkus field satisfy the perfectly conducting boundary condition, with $\B\bcdot\vec{n}=0$ at $\partial\mathcal{V}$, the solutions cannot be associated \change{to}{with} the SV at the CMB and a \add{more} suitable \add{background} magnetic field needs to be introduced.

\subsection{Numerical Calculation of the Modes}

The velocity and magnetic field perturbations are assumed to be periodic \add{in time}, i.e.
\begin{subequations}
\begin{linenomath*}
\begin{align}
  \tilde\vel(\pos,t) &= \tilde\vel(\pos)\exp(i\omega t),\\
  \tilde\B(\pos,t) &= \tilde\B(\pos)\exp(i\omega t).
\end{align}
\end{linenomath*}
\end{subequations}
Enumerating the \add{QG} velocity basis $\tilde\vel_i$, with $i=1,...,N_2$ and magnetic field basis $\tilde\B_i$, with $i=1,...,N_3$,
the projections \eqref{eq:varapproach} discretize to

\begin{linenomath*}
 \begin{equation}
    \mathrm{i} \omega\vec{M}\vec{x}=\vec{D}\vec{x}, \label{eq:gep}
\end{equation}
\end{linenomath*}
with $\vec{x}=(\hat\alpha_j,\zeta_j)\in \Cn^{N_2+N_3}$, and $\vec{M},\vec{D}\in \R^{N_2+N_3\times N_2+N_3}$ of the form
\begin{linenomath*}
\begin{align}
    \vec{M}=\begin{pmatrix}
    U_{ij} & 0\\
    0 & B_{ij}
    \end{pmatrix}, && \vec{D}=\begin{pmatrix}
    C_{ij} & L_{ij}\\
    V_{ij} & 0
    \end{pmatrix}.
\end{align}
\end{linenomath*}
Here, the coefficient matrices $U_{ij},C_{ij}\in\R^{N_2\times N_2}$, $L_{ij}\in\R^{N_2\times N_3}$, $B_{ij}\in\R^{N_3\times N_3}$ and $V_{ij}\in\R^{N_3\times N_2}$ correspond to the inertial \change{force}{acceleration}, Coriolis force, Lorentz force, \change{magnetic inertia force}{time change of magnetic field} and magnetic \change{advection}{induction} respectively.
This form is referred to as a generalized eigen problem solvable for eigen pairs $(\omega_k,\vec{x}_k)$.
A large part of $\vec{M}$ and $\vec{D}$ is zero, which allows us to use large polynomial degrees whilst keeping computational efforts small. 
To avoid numerical inaccuracies we use quadruple precision numbers (see supplementary material for more details). 
Our model extends the code available at \url{https://github.com/fgerick/Mire.jl} by the magnetic field basis introduced in \ref{sec:magbasis}.

\section{Results}
We choose a poloidal magnetic field of low polynomial degree, given by
\begin{linenomath*}
 \begin{equation}
 \begin{split}
  \B_0 =& \curl\curl (P_{100}+P_{110})\pos +\nabla\left( \alpha_{100}\Phi^i_{100}+\alpha_{110}\Phi^i_{110}\right)\\
  =& \frac{1}{c}\begin{pmatrix} x^2 + xz + 2y^2 + 2z^2 - 5/3\\ -xy + yz\\ -2x^2 - xz - 2y^2 - z^2 + 5/3 \end{pmatrix},\end{split}\label{eq:b0}
\end{equation}
\end{linenomath*}
with the constant\change{ $c=5\sqrt{\frac{184}{4725}\pi}$}{ $c=5\sqrt{\frac{368}{4725}}$}, so that $\vintin{\B_0\bcdot\B_0} = 1$.  
This field has a non-zero mean radial magnetic field at the equator. Its mean squared radial Alfv\'en velocity profile is given by
\begin{linenomath*}
 \begin{equation}
  v_A^2(s) = -\frac{189}{1840}s^4-\frac{42}{115}s^2+\frac{1281}{1840},\label{eq:va2}
\end{equation}
\end{linenomath*}
which decays slowly enough to ensure that TM are captured with low polynomial degrees.
\add{The Lehnert number is chosen to be $\Le=10^{-4}$, corresponding to a mean magnetic field strength of about 3 mT within Earth's core} \cite{gillet_fast_2010}.

\subsection{Properties of the Mode Spectrum}

\begin{figure}
    \centering
    \includegraphics[width=\textwidth]{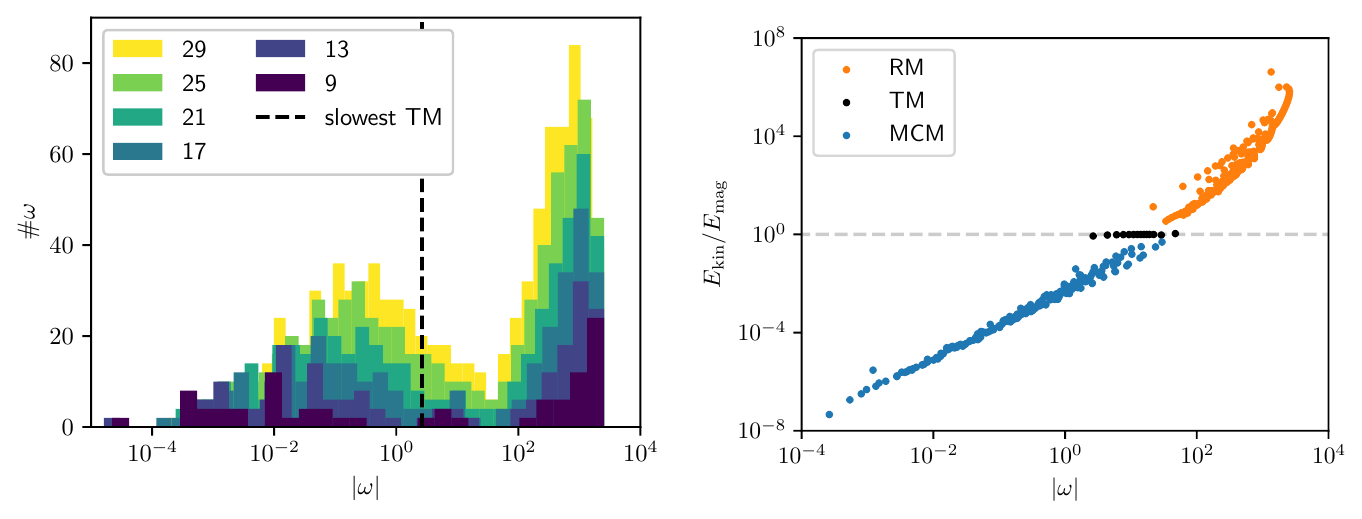}
    
    \caption{\add{An overlap of the} density of eigen solution spectrum\add{s} at different truncation\remove{s} $9\leq N\leq 29$ and $\Le=10^{-4}$ (left). Ratio of kinetic to magnetic energy \add{at $N=29$} (right). }
    \label{fig:density}
\end{figure}

The density of the eigen solution spectrum is shown in Figure \ref{fig:density} (left) for different degrees of truncation up to $N=29$.
The band limitation of the spectrum is easily seen on the fast end of the spectrum with a similar upper end of frequencies for all truncation degrees.
The fastest mode in the spectrum approximately corresponds to $|\omega^+_{1,5}| \approx 0.26\Le^{-1}$ (or $0.26\Omega$) instead of $|\omega|<2\Omega$ for inertial modes \cite{greenspan_theory_1968}.
For MCM the slowest frequency is affected by the truncation. This is due to the fact that the convergence of magnetic modes depends on the truncation, as the Lorentz force alters the polynomial degree of the modes, unlike the Coriolis operator \cite{ivers_enumeration_2015}.
For a truncation $N\leq 13$ TM are separated from the MCM and the fast modes. 
At larger $N$ some MCM are present also in the frequency range of TM.
At low truncation the classification of MCM, RM and TM is straightforward by the difference in frequencies.
At higher truncation we \change{differentiate}{classify} the modes by their kinetic and magnetic energies.

The kinetic and magnetic energies are respectively given by
\begin{linenomath*}
\begin{align}
    E_\mathrm{kin} &= \frac{1}{2}\vintin{\vel\bcdot\vel},\\
    E_\mathrm{mag} &= \frac{1}{2}\vintin{\B\bcdot\B}.
\end{align}
\end{linenomath*}

At the degree of $N=29$ both RM and MCM reach frequencies around the TM frequency range, but their energy ratio $\ekmr$ is still different \change{to}{from} unity (see Figure \ref{fig:density}, right).
At the considered polynomial degrees more MCM have periods comparable to those of TM than RM. This bias can be explained by the periods of Malkus modes as a function of \change{$N$}{$n$} and $m$ \cite<compare middle Figure 2 in>[]{labbe_magnetostrophic_2015}. 
For RM the periods only decrease towards unity as a function of \change{$N$}{$n$}, not of $m$ (unless $m\gg 1$). 
For MCM both an increase in $m$ and \change{$N$}{$n$} leads to an increase of the mode period towards unity.
This explains why at a certain truncation level the fastest MCM is closer to the periods of TM than the slowest RM.
Compared to the Malkus field, MCM (RM) spread out to higher (lower) frequencies and higher (lower) energy ratio. 

\subsection{Torsional Modes}

\begin{figure}
    \centering
    \includegraphics[width=0.8\textwidth]{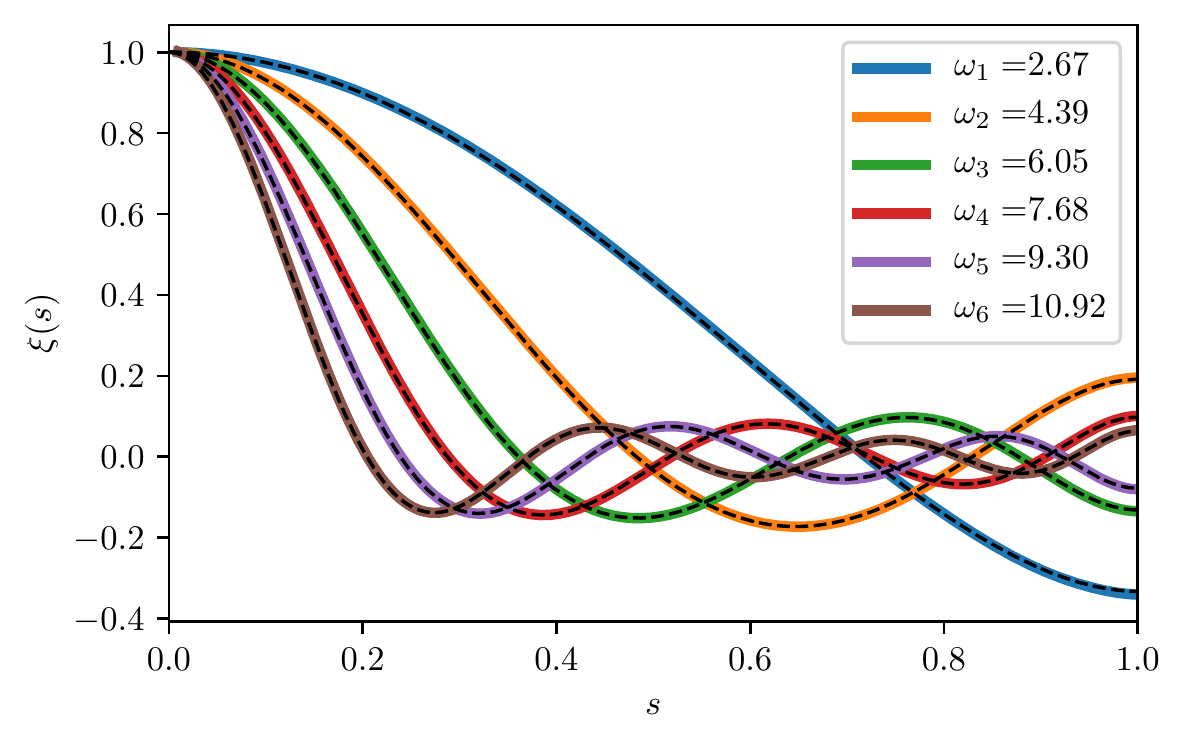}
    \caption{The \change{four}{six} largest scale TM of the hybrid model at $N=29$ and $\Le=10^{-4}$  (solid colors) and of the \change{1D}{1-D} equation (dashed black).}
    \label{fig:tm1}
\end{figure}

Given the profile of $v_A^2(s)$ we compute the TM by integrating \eqref{eq:1dtm} using finite differencing.
The \change{1D}{1-D} equation implies that $\partial \xi/ \partial s = 0$ at $s=1$ and it is automatically satisfied in our solver, if $v_A^2(1)\neq 0$.
The selection of TM in the spectrum of modes of the hybrid solution is done by considering the frequency range indicated by the \change{1D}{1-D} solutions and by a unit ratio of kinetic to magnetic energy of the eigen solutions.
The comparison between the \change{1D}{1-D} solutions (dashed black) and the hybrid model (solid colors) is shown in Figure \ref{fig:tm1}.
The six largest scale TM calculated by the hybrid model are in excellent agreement with the \change{1D}{1-D} solutions. 
The frequencies obtained from the two models have a relative difference \add{of} $\mathcal{O}(10^{-3})$.
We see that for both models $\partial \xi / \partial s =0$ at $s=1$, as expected.
For the hybrid model the resolution of this boundary condition depends on the \change{complexity}{radial wave number} of the TM in $s$, the \change{complexity}{spatial heterogeneity} of $\B_0$ and the polynomial truncation.
At a polynomial truncation of $N=29$ at least the six largest scale TM are well resolved by the basis.
The spatial structure of the magnetic field component of the TM depends only on the structure of $\B_0$ and the complexity of the TM in $s$ \cite<not presented here, see e.g.>[]{cox_observational_2016}.

\subsection{Slightly Modified Rossby Modes}

\begin{figure}
\includegraphics[width=\textwidth]{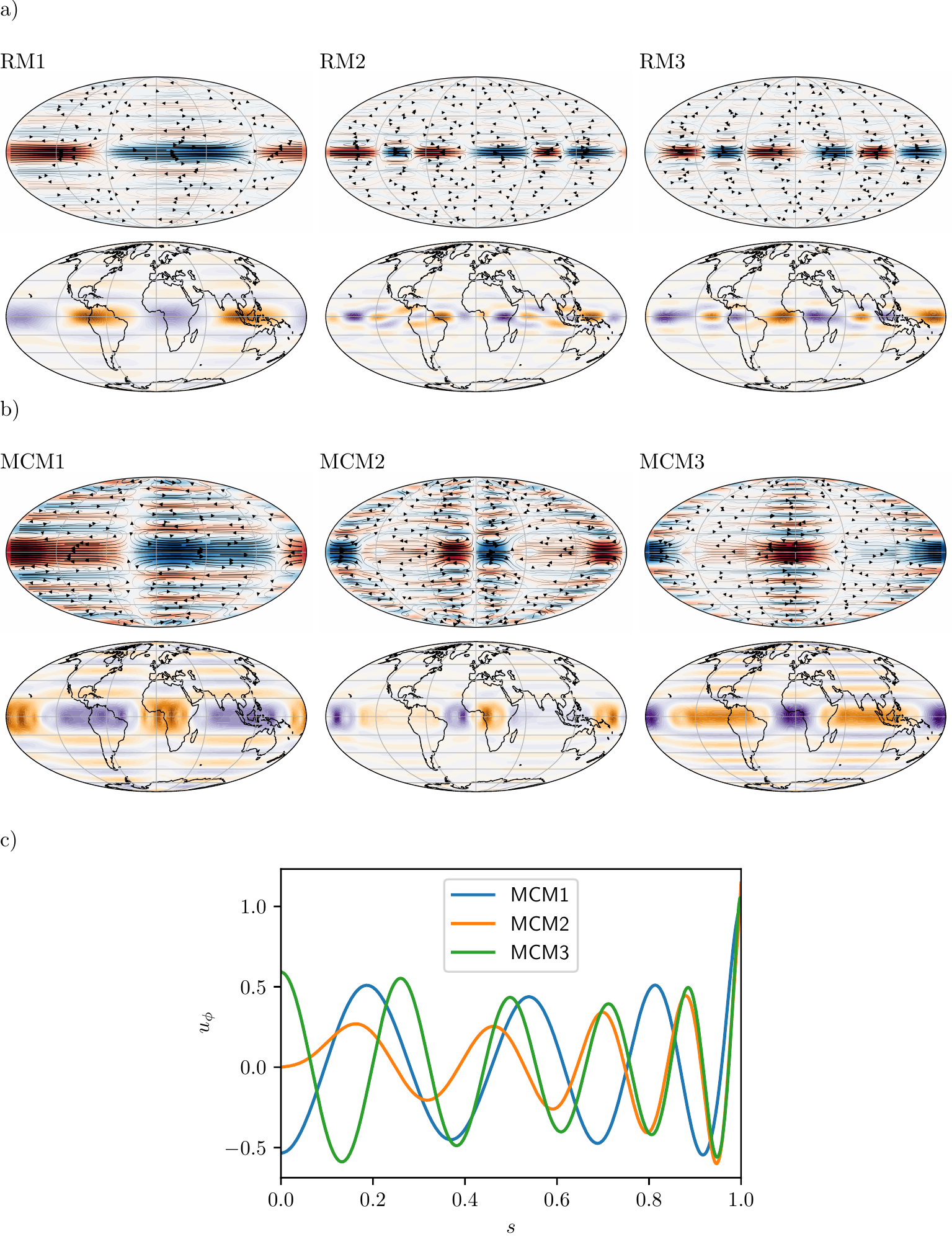}
    %   a) \hfill
    
    % \vspace{0.5cm}
    % \begin{minipage}[c]{0.32\textwidth}

    %     RM1
        
    %     \includegraphics[width=\linewidth]{rm_1_N35.png}
        
    %     \includegraphics[width=\linewidth]{rm_sv_1_N35.pdf}
   
    % \end{minipage}
    % \begin{minipage}[c]{0.32\textwidth}
    %     RM2
        
    %     \includegraphics[width=\linewidth]{rm_2_N35.png}
        
    %     \includegraphics[width=\linewidth]{rm_sv_2_N35.pdf}

    % \end{minipage}    \begin{minipage}[c]{0.32\textwidth}
    %  RM3

    %     \includegraphics[width=\linewidth]{rm_3_N35.png}
        
    %     \includegraphics[width=\linewidth]{rm_sv_3_N35.pdf}
    % \end{minipage}
   
    %   b) \hfill
    
    % \vspace{0.5cm}
    % \begin{minipage}[c]{0.32\textwidth}
    %       MCM1

    %     \includegraphics[width=\linewidth]{mcm_1_N35.png}
        
    %     \includegraphics[width=\linewidth]{mcm_sv_1_N35.pdf}
    
    % \end{minipage}
    % \begin{minipage}[c]{0.32\textwidth}
    %      MCM2

    %     \includegraphics[width=\linewidth]{mcm_2_N35.png}
        
    %     \includegraphics[width=\linewidth]{mcm_sv_2_N35.pdf}
    
    % \end{minipage}    \begin{minipage}[c]{0.32\textwidth}
    % MCM3

    %     \includegraphics[width=\linewidth]{mcm_3_N35.png}
        
    %     \includegraphics[width=\linewidth]{mcm_sv_3_N35.pdf}
    % \end{minipage}

    % \vspace{0.5cm}
    % c) \hfill
    
    % % \vspace{0.5cm}
    % \centering
    % \includegraphics[width=0.6\textwidth]{mcm_uphi1d_N35.pdf}
  \caption{\change{a) Surface flows (left column) and associated radial magnetic field perturbation at the surface (right column) of the three fastest converged RM. b) Surface flows (left column) and associated radial magnetic field perturbation at the surface (right column) of three converged MCM in the TM frequency range. Colors indicate the azimuthal velocity magnitude with blue being prograde and red being retrograde and inward (blue) and outward (orange) magnetic flux. The modes are computed for a maximum polynomial degree $N = 35$.}{a) Core surface flows (top row) and associated radial magnetic field perturbation at the surface (bottom row) of the three fastest, converged, RM (RM1--3). b) Core surface flows (top row) and associated radial magnetic field perturbation at the surface (bottom row) of three selected MCM (MCM1--3) in the TM frequency range. Colors indicate the azimuthal velocity magnitude with blue being prograde and red being retrograde and inward (blue) and outward (orange) magnetic flux. The arrows indicate tracers of the surface velocity. c) Cylindrical-radial profile of the azimuthal velocity in the equatorial plane at ${\phi=0}$ of MCM1--3. The modes are computed for a maximum polynomial degree ${N=35}$. The frequencies and periods of the modes are given in Table }\ref{tab:freqs}.}
  \label{fig:qginertialsurf} 
%   \caption{a) \add{Core} surface flows \change{(left column)}{(top row)} and associated radial magnetic field perturbation at the surface \change{(right column)}{(bottom row)} of the three fastest converged RM. b) \add{Core} surface flows \change{(left column)}{(top row)} and associated radial magnetic field perturbation at the surface \change{(right column)}{(bottom row)} magnitude with blue being prograde and red being retrograde and inward (blue) and outward (orange) magnetic flux. \add{The arrows indicate tracers of the surface velocity. c) Azimuthal velocity in the equatorial plane of three converged MCM.} The modes are computed for a maximum polynomial degree $N=35$.}
%   \label{fig:qginertialsurf}
\end{figure}

\add{For the range of polynomial degrees studied here,} RM are only slightly influenced by the presence of the magnetic field. 
Their spatial structure and frequency remain comparable to that of the RM in the purely HD case.
We compared the frequencies of some of the largest scale RM to the frequencies of the RM when including magnetic forces with the Malkus field \add{$\B_{0,M}=s\vec{1}_\phi$} and \remove{$\B_0$}\eqref{eq:b0}.
The \add{relative} differences between the frequencies of the three models are below \change{0.03}{$10^{-5}$} for all modes up to a truncation of $N=29$. \remove{This corresponds to a relative difference smaller than $10^{-5}$.} 
\change{This difference increases with increasingly large complexity of the modes under consideration}{This difference can increase at $N>29$}, as may be anticipated by the dispersion relation \eqref{eq:malkus}. 
The spatial structure also remains mostly unchanged and their phase velocity is prograde, as observed in the HD case.
We show the velocity and the radial magnetic field at the surface of three RM in Figure \ref{fig:qginertialsurf}a.
The slowest modes are associated \change{to}{with} $m=1$ and increasingly large $N$ (see the top mode). 
This can be seen also by a careful analysis of the approximated dispersion relation \eqref{eq:zhangrossby}.
We found even slower RM in our model, but we display three modes that are well converged from degree $N=29$ to $N=35$ (see details in supplementary material).
In both the HD and MHD case \change{a focusing}{an increase} of the velocity amplitudes near the equator is observed \cite<compare with RM in>[]{kloss_timedependent_2019}.
Their ratio of kinetic to magnetic energy is $\mathcal{O}(10)$, suggesting that their surface magnetic field perturbation \change{is significant}{might be observable in the future}.

\subsection{Magneto-Coriolis Modes \label{subsec:mcm}}

MCM are strongly influenced by the background magnetic field.
They are not easily compared between different magnetic fields, e.g. between the idealized Malkus field and the magnetic field $\B_0$.
Instead we focus on the MCM of \change{large}{relatively high} frequency, that are of particular interest here.
From Figure \ref{fig:density} we see that some MCM evolve on time scales similar to those of TM. 
We select three MCM \add{(MCM1--3)} with dimensionalized periods of a few years (see exact figures in Table \ref{tab:freqs}), that show a polynomial complexity below that of the truncation degree and a converged structure (see supplementary material).
The spatial structure of these selected MCM is presented in Figure \ref{fig:qginertialsurf}b, showing that they have a large complexity along the cylindrical radius and a relatively small azimuthal wave number. 
\add{The short length scale in cylindrical radius is even more evident in the radial profile of the azimuthal velocity in the equatorial plane, shown in Figure} \ref{fig:qginertialsurf}c.
\change{Similar}{Similarly} to the slow RM\add{,} the fast MCM \change{focus}{concentrate} their kinetic and magnetic energy near the \change{equatorial region}{equator}.
All MCM observed here travel retrograde, compared to the prograde direction of RM, as predicted by \citeA{hide_free_1966}.
For an azimuthal wave number $m=2$ in the equatorial band, the magnetic field perturbations of the three displayed MCM have a phase velocity of $\omega m^{-1} u_A \approx 680-2400$ km/yr. \add{However, we highlight that other wave numbers contribute to each mode with different phase velocities.}

\begin{table}
    \centering
    \caption{Non-dimensional frequency $\omega$, dimensionalized period $T$ in years\add{ (for $|\B_0|=3$ mT),} and ratio of kinematic to magnetic energy $\ekmr$ of the \change{6}{six} slowest TM (displayed in Figure \ref{fig:tm1}), the three fastest RM and MCM (displayed in Figure \ref{fig:qginertialsurf}). }\label{tab:freqs}
\begin{tabular}{cccc}
Type & $\omega$ & $T [\mathrm{yr}]$ & $E_\mathrm{kin}/E_\mathrm{mag}$\\
\hline\\
TM & 2.67 & 10.3 & 0.86 \\
TM & 4.39 & 6.2 & 0.94 \\
TM & 6.05 & 4.5 & 0.97 \\
TM & 7.68 & 3.6 & 0.98 \\
TM & 9.30 & 2.9 & 0.99 \\
TM & 10.92 & 2.5 & 0.99 \\
MCM1 & 1.72 & 15.91 & 0.02 \\ 
MCM2 & 2.66 & 10.28 & 0.03 \\ 
MCM3 & 5.95 & 4.60 & 0.13 \\ 
RM1 & 112.25 & 0.24 & 49.57 \\ 
RM2 & 118.09 & 0.23 & 12.26 \\ 
RM3 & 136.79 & 0.20 & 23.03
\end{tabular}
\end{table}

\section{Discussion}

We have shown, for $\Le= 10^{-4}$ and a mean magnetic field strength in the core interior of about $3$ mT, that changes of the magnetic field on periods as short as a few years could be explained by MCM.
Since the periods of these fast MCM are only a few years they may be associated \change{to}{with} the periodic secular acceleration impulses inferred from recent satellite observations \cite{chulliat_2015,kloss_timedependent_2019,chi-duran_2020}.
 These observations have been interpreted as the signature of MCM in the presence of a strong azimuthal magnetic field \cite{hori_slow_2015} or of MAC waves in a stratified layer at the top of the core \cite{knezek_influence_2018,buffett_equatorially_2019}. 
 We find there is no need to introduce a magnetic field stronger than inferred from TM or a stratified layer to account for fast wave propagation in the equatorial region of the core. 
 A key result of our study is the presence of large horizontal scales of $B_r$ at the core surface next to the equator associated with MCM, while the mode structure itself remains \change{complex}{small scale} in the cylindrical radial direction. 
 Such large magnetic features near the equator should be captured by satellite observations.
 Meanwhile, \citeA{aubert_geomagnetic_2019} have linked the secular variation impulses to so-called QG Alfv\'en waves arising near strongly heterogeneous magnetic fields of buoyant plumes in their numerical simulations.
Whether or not our fast MCM are in agreement with their explanation remains to be investigated.
\add{Our model could be used to invert geomagnetic observations for such a possible excitation mechanism, as described by} \citeA{buffett_inversion_2009}\add{ for TM.}

Fast MCM show \change{focusing}{a concentration of energy} \change{in}{near} the \change{equatorial region}{equator}, similar to the slowest RM.
Equatorially trapped waves have been much discussed either from observations \cite{chulliat_2015} or from physical models \cite{bergman_magnetic_1993}. The surface core flow calculations of \citeA{gillet_reduced_2019} also show the largest core flow acceleration pattern in an equatorial belt below $10^\circ$ of latitude. \change{Focusing}{Concentration of energy} of the modes could be favored by the weaker intensity of $\B_0$ in the equatorial region, which \citeA{ knezek_influence_2018, buffett_equatorially_2019} also found to be important for the focusing of MAC waves in the equatorial region of a stratified layer.
\remove{Irrespective of the magnetic field geometry, the increasingly steep slope of the CMB towards the equator suffices to produce equatorial mode focusing in our model.} 
A systematic study over a wider range of magnetic field geometries would be needed to make this statement quantitative.

\citeA{bergman_magnetic_1993} and \citeA{buffett_equatorially_2019} have shown that equatorially trapped MAC modes are strongly affected by damping.
We haven't included diffusion in our study and an investigation into how a diffusive layer at the top of the core may influence the observed fast MCM is necessary, even though they are of large spatial scale at the equator.
The new basis presented here potentially allows us to include magnetic diffusion, at a substantial computational cost. 

Previously, dynamics in the bulk of the core have been linked to inverted surface flows, but not directly to the observed changes in the magnetic field.
Being able to associate at once MCM, as well as TM, to magnetic field changes occurring with \remove{periods of }periods of 10 years or less and yet with large horizontal scale in an equatorial band at the CMB opens new perspectives for data assimilation and analyses of the dynamics occurring in the Earth's outer core.

\bibliography{biblio}

\appendix

\section{Derivation of Magnetic Field Basis\label{a:magbasis}}

Let us write the current density $\vec{j} = \curl\B$ in the toroidal-poloidal expansion, so that

\begin{linenomath*}
 \begin{equation}
    \vec{j} = \curl Q\pos + \curl\curl S\pos.
\end{equation}
\end{linenomath*}
We can also write the magnetic field in the toroidal-poloidal expansion, with
\begin{linenomath*}
 \begin{equation}
    \vec{B} = \curl T\pos + \curl\curl P\pos.
\end{equation}
\end{linenomath*}
It follows \cite{backus_foundations_1996}\remove{,} that
\begin{linenomath*}
\begin{align}
    S&=T,\\
    \nabla^2P &= -Q.\label{eq:poisson}
\end{align}
\end{linenomath*}

We construct the toroidal and poloidal scalars for the basis of $\vec{j}$ following the velocity basis introduced by \citeA{ivers_enumeration_2015}, so that
\begin{linenomath*}
\begin{align}
    Q_{lmn} &= r^{2n}R_l^m,\\
    S_{lmn} &= (1-r^2)r^{2n}R_l^m,
\end{align}
\end{linenomath*}
with $R_l^m = r^lY_l^m(\theta,\phi)$ the solid spherical harmonics. We have $|m|\leq l$ and $l \in [1,N]$, $n\in[0,(N-l)/2\big\rfloor$ for the toroidal basis and $l \in [0,N-1]$, $n\in[0,(N+1-l)/2-1\big\rfloor$ for the poloidal basis.
These $N_3=\frac{1}{6}N(N+1)(2N+7)$ elements form a complete basis for the current density in the set of polynomial vector fields of degree $N$ in the volume $\mathcal{V}$ \cite{ivers_enumeration_2015}.
Then, the toroidal part of $\B$ is directly given by
\begin{linenomath*}
 \begin{equation}
    \B_{t,lmn} = \curl S_{lmn} \pos. \label{eq:btor_a}
\end{equation}
\end{linenomath*}

For the poloidal part
\begin{linenomath*}
 \begin{equation}
    \B_{p,lmn} = \curl\curl P_{lmn}\pos,
\end{equation}
\end{linenomath*}
we need to solve the Poisson equation
\begin{linenomath*}
 \begin{equation}
    \nabla^2 P_{lmn} = -Q_{lmn}.
\end{equation}
\end{linenomath*}
We can use a slightly modified version of equation (3.1.9) in \citeA{backus_foundations_1996}
\begin{linenomath*}
 \begin{equation}
    \nabla^2 \left( r^{2(n+1)}\frac{R_l^m}{2(n+1)(2(l+n)+3)} \right)=r^{2n}R_l^m,
\end{equation}
\end{linenomath*}
to find that
\begin{linenomath*}
 \begin{equation}
    P_{lmn} = -r^{2(n+1)}\frac{R_l^m}{2(n+1)(2(l+n)+3)}.
\end{equation}
\end{linenomath*}

It remains to ensure that the internal magnetic field can be matched to a potential field at the boundary.
The poloidal component has to satisfy
\begin{linenomath*}
 \begin{equation}
    \nabla \Phi^i + \B_p = \nabla \Phi^e\quad\mathrm{at}\, \partial\mathcal{V}, \label{eq:vive}
\end{equation}
\end{linenomath*}
with $\Phi^i$ and $\Phi^e$ the interior and exterior potential field, respectively. The exterior potential field must vanish at infinity, if the source of the magnetic field lies within the interior. 
We can solve for $\Phi^i$ and $\Phi^e$ by considering the radial component of \eqref{eq:vive} and the horizontal component of \eqref{eq:vive}
\begin{subequations}
\begin{linenomath*}
\begin{align}
    \partial_r \Phi^i - \partial_r\Phi^e &= -B_{p,r},\\
    \nabla_H\Phi^i - \nabla_H \Phi^e &= -\B_{p,H}
\end{align}
\end{linenomath*}
\end{subequations}
Using the properties of the spherical harmonics, we find that
\begin{linenomath*}
 \begin{equation}
B_{p,r} = \frac{1}{r} L^2 P , \qquad \B_{p,H} = \nabla_H \left ( \frac{\partial}{\partial r}(rP) \right ),
\end{equation}
\end{linenomath*}
\add{with $L^2 P=\partial_r(r^2 \partial P/\partial r)-r^2\nabla^2P$} and the system simplifies to
\begin{subequations}
\begin{linenomath*}
\begin{align}
    \partial_r \Phi^i - \partial_r\Phi^e &= -L^2 P,\\
    \Phi^i - \Phi^e &= -\partial_r(rP).
\end{align}
\end{linenomath*}
\end{subequations}
Since this is linearly independent \change{in $l,m,n$}{of $l$, $m$, and $n$} we can consider this system for each $\B_{p,lmn}$ individually, so that
\begin{subequations}
\begin{linenomath*}
\begin{align}
    \partial_r \Phi^i_{lmn} - \partial_r\Phi^e_{lmn} &= -L^2 P_{lmn},\\
    \Phi^i_{lmn} - \Phi^e_{lmn} &= -\partial_r(rP_{lmn}).
\end{align}
\end{linenomath*}
\end{subequations}
Since $L^2Y_l^m = l(l+1)Y^l_m$ and
\begin{linenomath*}
\begin{equation}
    \partial_r(rP_{lmn}) = \frac{2(n+1)+l+1}{2(n+1)(2(l+n)+3)}r^{2(n+1)+l}Y_l^m,
\end{equation}
\end{linenomath*}
we need to search for potentials of the form
\begin{subequations}
\begin{linenomath*}
\begin{align}
    \Phi^i_{lmn} &= \alpha_{lmn}r^lY_l^m,\\
    \Phi^e_{lmn} &= \beta_{lmn}r^{-(l+1)}Y_l^m.
\end{align}    
\end{linenomath*}
\end{subequations}
The system to be solved for each $l,n$ is then
\begin{linenomath*}
\begin{equation}
    \begin{pmatrix}l&l+1\\1&-1\end{pmatrix}\begin{pmatrix}\alpha_{lmn}\\ \beta_{lmn}\end{pmatrix} = -\frac{1}{\alpha}\begin{pmatrix}l(l+1)\\2n+l+3\end{pmatrix}, \label{eq:vivels}
\end{equation}
\end{linenomath*}
with $\alpha=2(n+1)(2(l+n)+3)$, for each $l,n$. The solutions to \eqref{eq:vivels} are
\begin{subequations}
\begin{linenomath*}
\begin{align}
    \alpha_{lmn} &= -\frac{(l + 1)}{(2l + 1)(2n + 2)},\\
    \beta_{lmn} &= \frac{l}{(2l+1)(2l+2n+3)},
\end{align}
\end{linenomath*}    
\end{subequations}
so that the poloidal basis vectors are given by
\begin{linenomath*}
 \begin{equation}
    \B_{p,lmn} = \curl\curl P_{lmn}\pos + \nabla \Phi^i_{lmn}.\label{eq:bpol_a}
\end{equation}
\end{linenomath*}

\citeA{li_optimal_2010} presented a similar basis to express the magnetic field for an insulating mantle, which involves slightly more complicated expressions for the poloidal and toroidal scalars with Jacobi polynomials.
We have not proven any weighted orthogonal inner products of the poloidal and toroidal scalars, but our basis vectors show the same orthogonality for the unweighted inner product between vectors of different harmonic order and degree as does the basis of \citeA{li_optimal_2010}. 
\add{Orthogonal bases, based on Jacobi polynomials and spherical harmonics, have been presented in} \citeA{chen_optimal_2018, li_taylor_2018}.
\add{These bases are desirable for reducing computational efforts, but without an orthogonal QG basis, no such computational advantage is given for the hybrid model presented here.}

\section{Spherical Harmonics in Cartesian Coordinates\label{a:csh}}

The unnormalized spherical harmonics are defined as
\begin{linenomath*}
 \begin{equation}
    \tilde Y_l^m(\theta,\phi) = e^{im\phi}P_l^m(\cos\theta),
\end{equation}
\end{linenomath*}
with the so-called associated Legendre functions
\begin{linenomath*}
 \begin{equation}
    P_l^m(x) = \frac{1}{2^l l!}\left(1-x^2\right)^{m/2} \frac{\partial^{m}}{\partial x^{m}}P_l(x)\label{eq:alp}
\end{equation}
\end{linenomath*}
with
\begin{linenomath*}
 \begin{equation}
    P_l(x) = \frac{\partial^{l}}{\partial x^{l}}\left(x^2-1\right)^l.
\end{equation}
\end{linenomath*}

The unit Cartesian coordinates are given by
\begin{linenomath*}
\begin{align}
    \hat x&=x/r=\cos\phi\sin\theta,\\
    \hat y&=y/r=\sin\phi\sin\theta,\\
    \hat z&=z/r=\cos\theta.
\end{align}
\end{linenomath*}
We can rewrite \eqref{eq:alp}, so that
\begin{linenomath*}
 \begin{equation}
    P_l^m(\cos\theta) = P_l^m(\hat z) = \frac{1}{2^l l!}\left(\sin\theta\right)^{m} \frac{\partial^{m}}{\partial \hat{z}^m}P_l(\hat{z}).
\end{equation}
\end{linenomath*}

Using the trigonometric identities
\begin{linenomath*}
\begin{align}
\sin(m\phi) &= \sum_{k\text{ odd}} (-1)^\frac{k-1}{2} \binom{m}{k}(\cos\phi)^{m-k} (\sin\phi)^k, \\
\cos(m\phi) &= \sum_{k\text{ even}} (-1)^\frac{k}{2} \binom{m}{k}(\cos\phi)^{m-k} (\sin\phi)^k,
\end{align}
\end{linenomath*}
we are able to rewrite the spherical harmonics in terms of the unit Cartesian coordinates, so that
\begin{linenomath*}
 \begin{equation}
\begin{split}
    \tilde Y_l^m(\theta,\phi) =& \tilde Y_l^m(\hat{x},\hat{y},\hat{z})\\
 =&\frac{1}{2^l l!} \left(\sum_{k\text{ odd}} (-1)^\frac{k-1}{2} \binom{m}{k} \hat{x}^{m-k}\hat{y}^k\right.\\
    &\left. + i\sum_{k\text{ even}} (-1)^\frac{k}{2} \binom{m}{k} \hat{x}^{m-k}\hat{y}^k \right) \frac{\partial^{m}}{\partial \hat{z}^m}P_l(\hat{z}).
    \end{split}
\end{equation}
\end{linenomath*}

The solid spherical harmonics are

\begin{linenomath*}
 \begin{equation}
    \tilde{R}_l^m(x,y,z) = r^l\tilde{Y}_l^m(\hat{x},\hat{y},\hat{z}),
\end{equation}
\end{linenomath*}
which is polynomial in $x$, $y$ and $z$.

\acknowledgments
The authors like to thank Henri-Claude Nataf and two anonymous reviewers for their help in improving this work.
FG was partly funded by Labex OSUG@2020 (ANR10 LABX56). 
Support is acknowledged from the European Space Agency through contract 4000127193/19/NL/IA. 
This work has been carried out with financial support from CNES (Centre National d'\'Etudes Spatiales, France).
JN was partly funded by SNF Grant \#200021\_185088.
Computations were performed on ETH Zurich's Euler cluster.
The data and the code to calculate it can be found at \url{https://dx.doi.org/10.5281/zenodo.4008396}.

\end{document}

%% ------------------------------------------------------------------------ %%
%
%  TITLE
%
%% ------------------------------------------------------------------------ %%

%\includegraphics{agu_pubart-white_reduced.eps}

\title{Supporting Information for "Fast Quasi-Geostrophic Magneto-Coriolis Modes in the Earth's core"}

\authors{F. Gerick\affil{1,2}, D. Jault\affil{1}, J. Noir\affil{2}}

\affiliation{1}{CNRS, ISTerre, University of Grenoble Alpes, Grenoble, France}
\affiliation{2}{Institute of Geophysics, ETH Zurich, Zurich, Switzerland}

\begin{article}

\section*{Solving the Generalized Eigen Problem\label{a:precision}}
\begin{figure}
    \centering
  \includegraphics[width=0.6\textwidth]{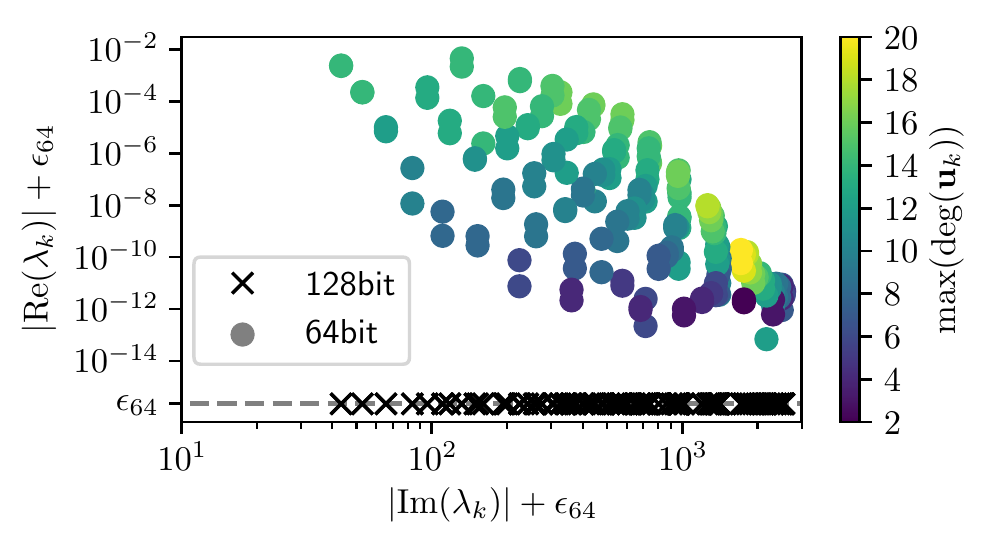}
  \caption{Comparison of eigen values to the Malkus problem using a 64bit accurate solver and a 128bit accurate solver for a truncation of $N=20$. We focus here on the frequency regime of the fast RM. The solutions of the 64bit solver are colored by the polynomial degree of their peak amplitude basis vector. The solutions to the 128bit solver are shown as orange crosses and lie on the line of $\mathrm{Re}(\lambda_k)+\epsilon_{64}\approx\epsilon_{64}\approx 2.2\times 10^{-16}$, the 64bit floating point error.}
  \label{fig:64bitvs128bit}
\end{figure}

The generalized eigen problem
\begin{equation}
    \mathrm{i} \omega\vec{M}\vec{x}=\vec{D}\vec{x}, \label{eq:gep}
\end{equation}
is transformed to a standard eigenvalue problem by inverting $\vec{M}$, so that
 \begin{equation}
 \lambda\vec{x}=\vec{A}\vec{x}, \label{eq:sep}
\end{equation}
where $\lambda=\mathrm{i}\omega$ and $\vec{A}=\vec{M}^{-1}\vec{D}$. In practice this inverse is not calculated explicitly, as $\vec{M}^{-1}$ is not necessarily sparse and its calculation costly.
Instead we factorize $\vec{M}$ to be able to solve the linear problem $\vec{A}\vec{x}=\vec{b}$ for $\vec{x}$.
The standard eigenvalue problem \eqref{eq:sep} can then be solved iteratively for the largest eigen pairs $(\lambda_k,\vec{x}_k)$ using the \change{implicitly restarted Arnoldi}{Krylov-Schur} method implemented in the Julia programming language.
The advantage of this implementation is an easy adaptation to higher accuracy floating point numbers.
In our model we observe an increase in the real part of the numerical eigen solutions for increasing polynomial degree due to numerical inaccuracies, demonstrated in Figure \ref{fig:64bitvs128bit} for the Malkus field.
We compensate this by using quadruple precision (128bit) floating point numbers throughout all of our computations, which are able to ensure stable enough solutions.
Due to the hybrid model a large portion of the eigen solutions are degenerate modes of zero frequency so that we choose to calculate only the non-zero modes.

Another difficulty, present for magnetic fields that are not linear in $x$,$y$ and $z$, is the convergence of the modes.
In this case spurious modes are always possible and they can perturb the physically relevant modes significantly.
We have thus to make sure that the calculated modes are converged.
To do so, first, the frequency shouldn't change significantly between different truncation degrees. 
Further, the velocity and magnetic field must also not change more than a given threshold. We impose that the mode calculated at a degree $N$, $\vel_{N}$, must be able to be matched to a mode calculated at a higher degree $N+2$, $\vel_{N+2}$, with a correlation
 \begin{equation}
    \frac{\vintin{\vel_{N}\bcdot\vel_{N+2}}}{\vintin{\vel_{N}\bcdot\vel_{N}}} > 1-\epsilon,
\end{equation}
with $\epsilon$ the allowed error.

\begin{figure} 
   \includegraphics[width=\textwidth]{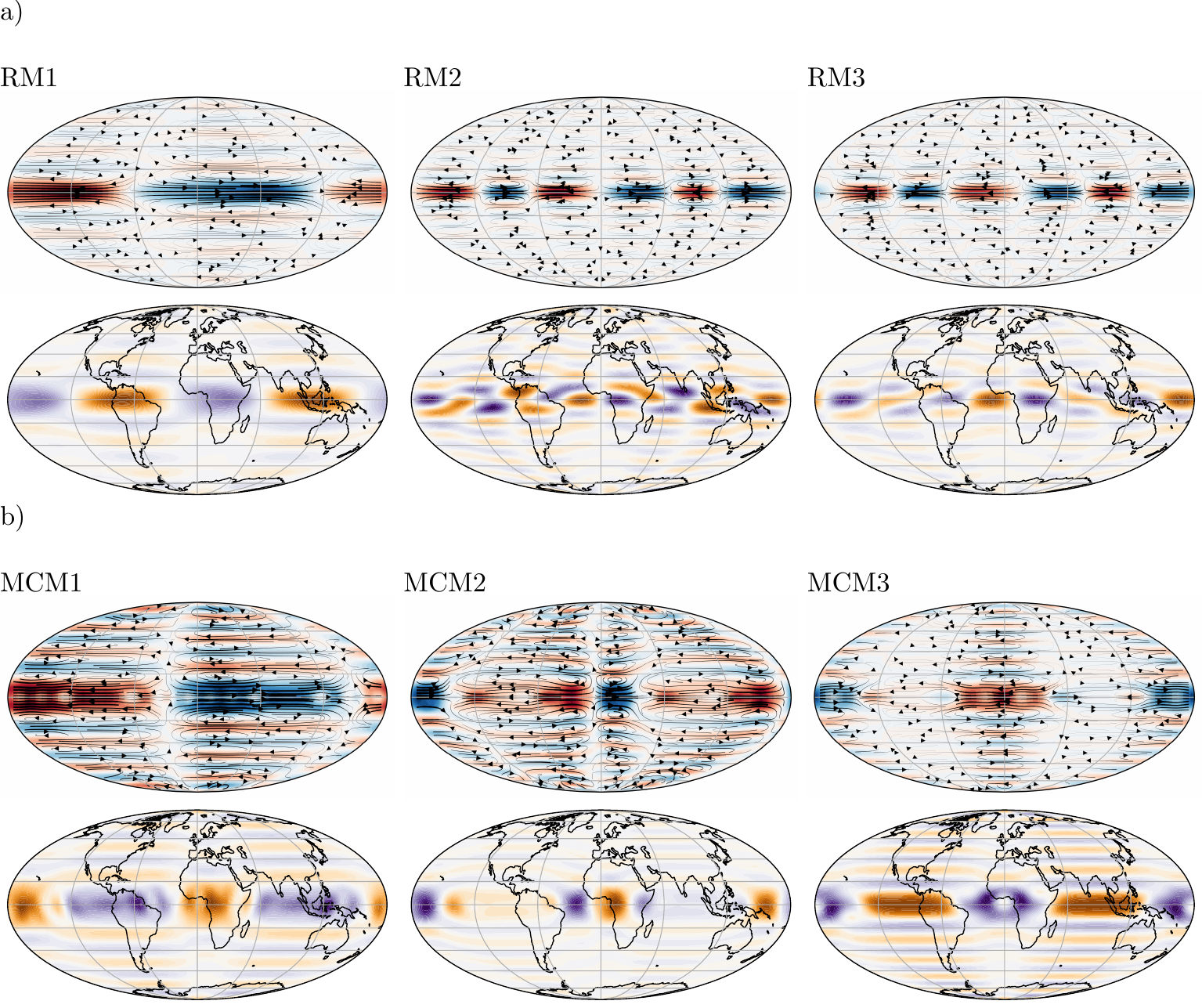} 
  \caption{Same as Figure 3 of the article with truncation degree $N=29$.}
  \label{fig:qginertialsurfN29}
\end{figure}

For the RM we choose $\epsilon =0.01$ and from a degree $N=29$ to $N=35$ the three modes presented in Figure 3a and \ref{fig:qginertialsurfN29}a are the fastest modes that satisfy this constraint.
For the MCM the threshold is lowered to $\epsilon = 0.05$. 
This way we ensure that for $N=29,31,33,35$ the modes keep their spatial structure approximately.
We reproduce Figure 3 of the article in Figure \ref{fig:qginertialsurfN29} to illustrate the difference.
Especially the equatorial region of the MCM is prone to the influence of spurious modes.
Nevertheless the broad structure, e.g. dominant wave number and equatorial focusing, are preserved throughout different degrees of truncation, making us confident that these modes are indeed sufficiently converged.
\end{article}
\clearpage

